\RequirePackage{fix-cm}
\documentclass[smallextended]{svjour3}       
\smartqed  
\usepackage{graphicx}
\usepackage{amsmath}
\usepackage{enumerate}

\begin{document}

\title{Gaussian Anisotropy In Strange Quark Stars}


\author{H. Panahi         \and
        R. Monadi         \and
        I. Eghdami
}


\institute{H. Panahi \at
              Department of Physics, University of Guilan, Rasht 41635-1914, Iran \\
              \email{t-panahi@guilan.ac.ir}
           \and
           R. Monadi \at
              Department of Physics, University of Guilan, Rasht 41635-1914, Iran\\
              \email{reza.monadi.90@gmail.com}
               \and
           I. Eghdami \at
              Department of Physics, University of Guilan, Rasht 41635-1914, Iran\\
              \email{issaeghdami@phd.guilan.ac.ir}
}

\date{Received: date / Accepted: date}

\maketitle

\begin{abstract}
In this paper for studying the anisotropic strange quark stars, we assume that the
radial pressure inside the anisotropic star is a superposition of pressure in an isotropic case plus a Gaussian perturbation term. Considering a proportionality between electric charge density and the density of matter, we
solve the TOV equation for different cases numerically. Our
results indicate that anisotropy increases the maximum mass
$M_{max}$ and also its corresponding radius $R$ for a typical strange
quark star. According to our calculations, an anisotropy amplitude of $A=3\times10^{33}Nm^{-2}$ with a standard deviation of $\sigma=3\times10^{3}m$ leads to a neutron star of  1.97$M_{\odot}$. Furthermore, electric charge not only increases the maximum mass and its corresponding
radius, but also raises up the anisotropy factor. We can see that the tangential
pressure $p_{t}$ and anisotropy factor $\Delta$ unlike the radial pressure $p_{r}$ have a maximum on the surface and this maximum increases by adding electric charge effect. However, we show that  anisotropy can be more effective than electric charge  in rasing maximum mass of strange quark stars.

\end{abstract}

\section{Introduction}
Strange quark stars are hypothetical type of compact exotic stars which are firstly speculated in Refs.~\cite{alcock,haensel1986,farhi}.
Describing the structure of stars, including mass
$(M)$ and radius ($R$) will be available by solving hydrostatic
equilibrium equations. In the case of compact stars, due to high
density, General Relativity (GR) dominates and Newtonian
hydrostatic equilibrium should be replaced by its GR counterpart.
Assuming a static and spherical symmetric geometry and an
isotropic matter, Einstein field equations lead to
Tolman$-$Oppenheimer$-$Volkoff equation (TOV) \cite{weinberg}.
The structure of compact stars has been investigated by
Refs.~\cite{Haensel,harko,Singh,{maharaj}} and
many others by solving TOV equation using a suitable equation of
state. Now in this paper, we deal with the strange quark matter
(SQM) which its EOS is well described by MIT Bag Model \cite{Haensel}.\\
nuclear matter in very high densities can be anisotropic and it is a great
motivation to study the effect of anisotropy in the structure of
relativistic stars \cite{harko,Ruderman}. Additionally, existing a solid core and a
strong magnetic field in neutron stars can be related to
anisotropy in the matter of star \cite{harko,Bailin}.
Solving anisotropic TOV equation requires a physically reasonable
assumption. For instance, in Ref.~\cite{harko} a specific density
profile $\rho(r)$ has been chosen and in Ref.~\cite{maharaj} a special
metric function $\Lambda(r)$ has been utilized to solve it
analytically. But in this paper, we try to apply a numerical
method for solving modified TOV equation by considering a
perturbation term. Initially, we consider an isotropic and free
charge matter and obtain a solution for radial pressure $p_r(r)$. Then in anisotropic
case we use the previous $p_r(r)$ as the unperturbed solution in order to study  $R$ and $M$ in this kind of stars.
It should be noted that in this work we are dealing with a non-rotating strange quark star but we can expect that rotation allow a larger maximum mass (about $40\%$) according to Ref. \cite{Haensel}.\\
This paper is organized as follows: In \S2 we first solve the TOV
equation for an isotropic matter and then compare it with a
charged and anisotropic one. Discussion and concluding remarks are
represented in \S3.

\section{TOV Equation}
\label{modified} For describing mass and radius of a self
gravitating configuration in relativistic term we have to consider
a suitable metric. The line element in interior of star, assuming
a static and spherically symmetric geometry can be written as:
\begin{equation}
ds^{2}=-c^{2}e^{2\phi}dt^{2}+e^{2\Lambda}dr^{2}+r^{2}d\Omega^{2},
\label{metric}
\end{equation}
where $c$ is the light speed while $\phi$ and $\Lambda$ are spherically symmetric metric functions.\\
In order to obtain a General Relativistic hydrostatic equilibrium,
we have to solve the Einstein equations:
\begin{equation}
\label{ein}
G_{\mu\nu}=\frac{8\pi G}{c^{4}}T_{\mu\nu},
\end{equation}
for the metric given in Eq. (\ref{metric}). Also we must choose a
reasonable energy-momentum tensor which satisfies the following
conservation law:
\begin{equation}
\nabla_{\nu}T^{\mu\nu}=0.
 \label{cons}
\end{equation}
Now in the next subsections we try to obtain the TOV equation for
two different cases.

\subsection[]{Uncharged Isotropic Matter}
First of all we consider an uncharged perfect fluid. The
energy-momentum tensor of a perfect fluid is given by \cite{camenzind}:

\begin{equation}
  T^{\mu}_{\nu} =
 \begin{pmatrix}
  -\rho c^{2} & 0 & 0 & 0 \\
  0 & p & 0 & 0 \\
   0 & 0 & p & 0 \\
   0 & 0 & 0 & p
  \end{pmatrix},
 \label{T1}
\end{equation}

where $p$ and $\rho$ are the pressure and mass density
respectively which are spherical symmetric \cite{harko}.
We also define the gravitational mass as \cite{glender}:
\begin{equation}
m(r)=\int_{0}^{r} 4\pi x^{2}\rho(x) dx.
 \label{mm}
\end{equation}
By using the above equation, the Einstein equations (\ref{ein})
and the conservation law (\ref{cons}) for metric (\ref{metric})
yield to hydrostatic equilibrium equation:

\begin{equation}
\frac{dp}{dr}=-(\rho c^{2}+p)\frac{4\pi Gp
r^{3}+mGc^{2}}{rc^{2}(rc^{2}-2mG)}.
 \label{TOV}
\end{equation}
Noticeably $m$, $p$, and $\rho$ are functions of radial coordinate (\textit{$r$}). Equation (\ref{mm}) together with Eq. (\ref{TOV}) is called TOV
equation which for its solvability it has to be supplemented by an
EOS \cite{Haensel-book}. In this work we consider the MIT Bag Model for
EOS of SQM as:
\begin{equation}
\frac{p}{c^{2}}=\alpha(\rho-\rho_{s}).
 \label{EOS}
\end{equation}
According to Ref.~\cite{negi} for SQM2 model
$\rho_{s}= 3.056\times10^{17}kg$ $m^{-3}$ and $\alpha=0.324$ .
Now we can solve TOV equation numerically, using physical boundary conditions \cite{camenzind}: 
\begin{subequations}
\begin{center}
\begin{align}
p(0) &= p_{c}\label{bond1}\\
m(0) &= 0\label{bond2}\\
p(R) &= 0\label{bond3}\\
m(R) &= M\label{bond4}
\end{align}
\end{center}
\end{subequations}
In addition, in the stellar interior, following conditions should be satisfied:
\begin{subequations}
\begin{center}
\begin{align}
 p > 0 \label{cond1}\\
 \frac{dp}{dr} < 0 \label{cond2}\\
 \sqrt{\frac{dp}{d\rho}} \leq c\label{cond3}\\
 \frac{dp}{d\rho}\geq 0\label{cond4}
\end{align}
\end{center}
\end{subequations}
Conditions (\ref{cond1}) and (\ref{cond2}) are trivial for
preserving hydrostatic equilibrium and (\ref{cond3}) refers to
causality inside the star which does not permit
 $\alpha > 1$ in Eq. (\ref{EOS})  \cite{Haensel}. Also the EOS should satisfy the microscopic stability
condition (\ref{cond4}), otherwise the star collapses spontaneously\cite{shapiro}. In the next section we will use the obtained pressure $p(r)$
as the unperturbed radial pressure for anisotropic case.

\subsection[]{Charged Anisotropic Matter}
In this section instead of a perfect fluid we are dealing with an anisotropic matter. It means that pressure in radial and tangential directions are not necessarily the same. With this assumption, which is predicted in very high density ranges\cite{harko}, the energy momentum tensor in CGS unit system will be \cite{maharaj,newmalheiro}:
\begin{equation}
  T^{\mu}_{\nu} =
 \begin{pmatrix}
  -(\rho c^{2}+\frac{E^{2}}{8\pi }) & 0 & 0 & 0 \\
  0 & p_r-\frac{E^{2}}{8\pi } & 0 & 0 \\
   0 & 0 & p_t+\frac{E^{2}}{8\pi } & 0 \\
   0 & 0 & 0 & p_t+\frac{E^{2}}{8\pi }
 \end{pmatrix},
 \label{TA}
\end{equation}
where $p_{t}$ and $p_{r}$ are tangential and radial pressure
respectively. The gravitational mass which is the total
contribution of the energy density takes the new form as
\cite{malheiro}:
\begin{equation}
m(r)=\int_{0}^{r} 4\pi x^{2} \left( \rho(x)+\frac{E(x)^{2}}{8\pi
c^{2}} \right) dx,
 \label{mmnew}
\end{equation}
where $E$ is the radial electric field, defined as:
\begin{equation}
E(r)=\frac{1}{r^{2}}\int_{0}^{r}4\pi x^{2} \rho_{ch}e^{\Lambda}
dx
 \label{Electric}
\end{equation}
and $\rho_{ch}$ is the charge density. Defining $\Delta= p_{t}-p_{r}$ as anisotropy factor
\cite{harko}, Einstein equations (\ref{ein}) and conservation
equation (\ref{cons}) lead to:

\begin{equation}
e^{-2\Lambda}=1-\frac{2mG}{rc^{2}},
\label{lambda}
\end{equation}

\begin{equation}
 \frac{dp_{r}}{dr}= \left(\frac{2}{r}\Delta -\rho
c^{2}-p_{r}\right) \frac{4\pi
Gr^{3}(p_{r}-\frac{E^{2}}{8\pi})+mGc^{2}}{rc^{2}(rc^{2}-2mG)}+E\rho_{ch}e^{\Lambda}.
\label{TOVA}
\end{equation}
It should be noted that if $\Delta>0$, in Eq. (\ref{TOVA}) the magnitude of radial pressure derivative decreases. This causes a softer falling of radial pressure which results higher maximum masses and their corresponding radii. Equations (\ref{mmnew}) and (\ref{TOVA}) represent the modified
form of TOV equation. It is easy to see that putting $E=0$ and
$\Delta=0$ in Eq. (\ref{TOVA}), simply return uncharged and isotropic cases. In an anisotropic
matter, MIT Bag Model is \cite{harko}:
\begin{equation}
 \frac{p_{r}}{c^{2}}=\alpha(\rho-\rho_{s}).
 \label{EOS1}
\end{equation}
We should notice that tangential pressure $p_{t}$  does not necessarily vanish on
the surface of star, in contrary to $p_r$\cite{harko,maharaj}. Solving modified TOV  equation requires some extra assumptions.
Assuming that anisotropy has a slight effect on $p_r$, we add a
perturbation to isotropic pressure profile to obtain $p_r$ as:
\begin{equation}
p_{r}(r)_{Anisotropic}=p(r)_{Isotropic}+\delta p(r).
 \label{assume}
\end{equation}
The following perturbation function has to satisfy these boundary
conditions:
\begin{subequations}
\begin{align}
p_{r}(0)=p_{c},\label{s1}\\
p_{r}(R)=0,\label{s2}\\
\frac{dp_{r}}{dr}|_{r=0}=0,\label{s3}\\
\frac{dp_{r}}{dr}|_{r=R}=0.\label{s4}
\end{align}
\end{subequations}
Regarding to Eq. \ref{assume}, $p(r)_{Isotropic}$ and $p_{r}(r)_{Anisotropic}$ have to satisfy mentioned boundary conditions. So  $\delta p(r)$  must meet those conditions too. Gaussian function is the simplest choice (but not the only one\footnote{For example Lorentzian
function could be another choice for $\delta p(r)$ which yields no
significant difference.}) for fulfilment of the conditions \ref{s1} through \ref{s4} by setting the position of Gaussian pressure profile at $\mu=\frac{2R}{3}$ in the following equation:
\begin{equation}
\delta p(r)=A \exp \left(  \frac{-(r-\mu)^{2}}{2\sigma^{2}}
\right).
\label{gaussian}
\end{equation}
One can see that this Gaussian perturbation can satisfy mentioned boundary conditions physically. We need $\delta$ and its derivative to be nearly zero at origin and also on the surface. In fact we can tune $A$ and $\sigma$ so that the perturbation to be so small at those points. According to Ref.~\cite{harko} the anisotropy can cause an increase in the radial pressure. Hence we have
chosen a positive and small enough value compared to $p_r$ for
parameter $A$.
Eventually studying charged case requires one more physically assumption. According to Ref.~\cite{malheiro} electrical charge density can be related to the
density of matter as follows:
\begin{equation}
\rho_{ch}=f\times\rho,
\label{ff}
\end{equation}
which is a physically reasonable assumption\cite{malheiro}. In this equation $f$ is the charge
fraction that can be considered $f\leq10^{-5} esu$ $g^{-1}$ to satisfy
causality, stability, and electrical neutrality conditions of stars \cite{glender}. Therefore in the most general case (charged anisotropic matter), there is 7 variables namely $m, p_{r}, p_{t}, \rho, \rho_{ch}, E, \Lambda  $ and 7 equations (\ref{mmnew}) -  (\ref{assume}) and (\ref{ff}). We have used Runge Kutta Fehlberg fourth-fifth order method (RKF45 Method) to solve the corresponding coupled ODE systems in each cases below:
\begin{enumerate}
  \item Isotropic uncharged ($\Delta=0, E=0$ ),
  \item Isotropic charged ($\Delta=0, E\neq0$),
  \item Anisotropic uncharged ($\Delta\neq0, E=0$),
  \item Anisotropic charged ($\Delta\neq0, E\neq0$).
\end{enumerate}

\section{Discussion}
In this paper unlike the analytical solutions for anisotropic matter such as Ref.~\cite{harko}
and Ref.~\cite{maharaj} we have used numerical result of isotropic solution and
a Gaussian perturbation rather than using a totally assumed function.
In the following subsections we will discuss the result of adding anisotropy and electric charge to TOV equations.
Our solution indicate that adding positive Gaussian anisotropy and electric charge
to TOV equation increase the maximum mass of neutron star and can predict more massive neutron stars in agreement with Ref.~\cite{Demorest}.

\subsection{Charge Effect}
\begin{figure}
\centering
\includegraphics[scale=0.4]{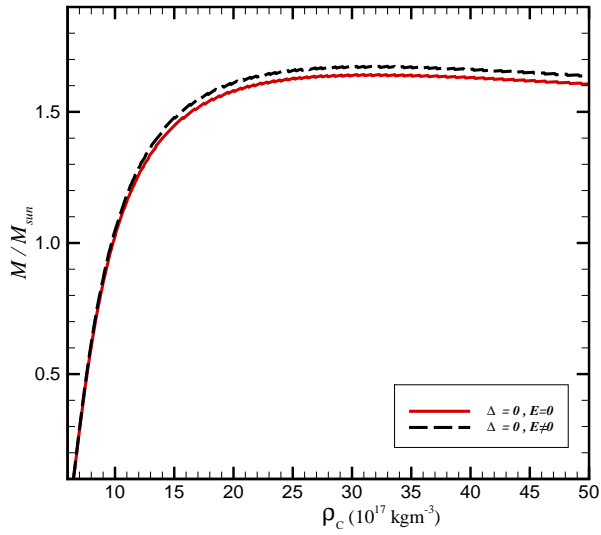} \\
\caption {Gravitational mass M versus central density $\rho_c$ in two
cases: Uncharged (solid curve) and charged isotropic (dashed
curve) matter. }
 \label{fig:MmaxE}
\end{figure}

\begin{figure}
\centering
\includegraphics[scale=0.4]{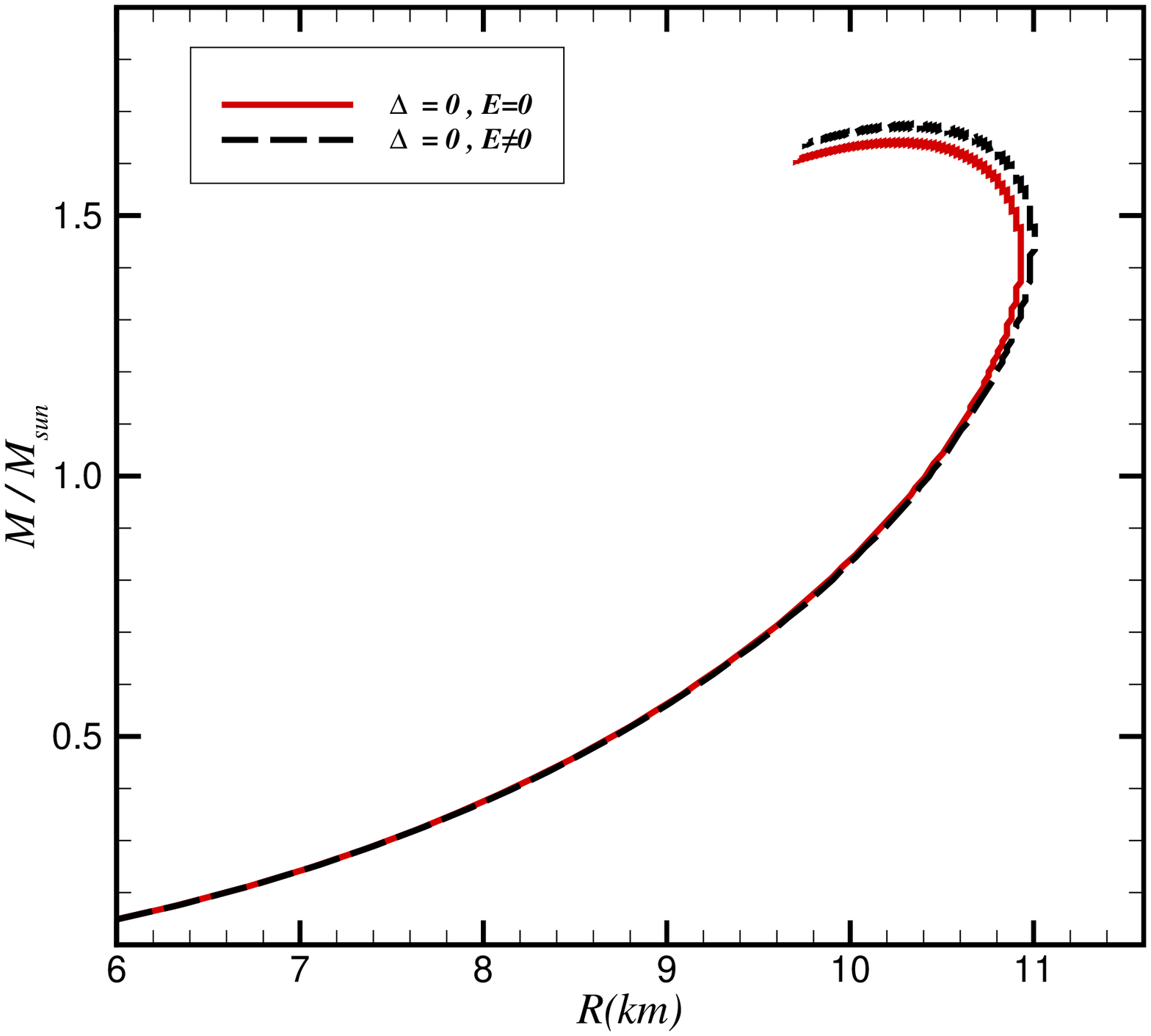} \\
\caption {Gravitational mass M versus radius of star R in two
cases: Uncharged isotropic (solid curve) and charged isotropic
(dashed curve) matter. }
 \label{fig:pichE}
\end{figure}

Solving modified TOV equation reveals that adding electric charge makes the radial pressure $p_r$ to be increased, as we can see in Fig. \ref{fig:p_rE}. But the sensitivity of tangential pressure $p_t$ to the electric charge is much more, so $\Delta$ raises up in charged case as is apparent from Fig. \ref{fig:deltaE}. This behavior is not only
the direct result of our speculation about $p_r$, but also comes
from the nature of TOV equation.\\
Fig. \ref{fig:MmaxE} and Fig. \ref{fig:pichE} refer to uncharged and charged isotropic
cases. One may expect adding electric charge to TOV equation causes an
increase in the maximum mass of strange quark star $M_{max}$ and
also in the corresponding radius $R$ for obtained $M_{max}$. Since
in GR gravitational mass of star and total energy are
proportional, adding charge density $\rho_{ch}$ increases the
total energy and therefore the above result makes sense.
\cite{glender}. However, electrical neutrality condition of stars does not allow a significant growth in maximum masses and their corresponding radii.

\subsection{Anisotropy Effect}
\begin{figure}
\centering
\includegraphics[scale=0.4]{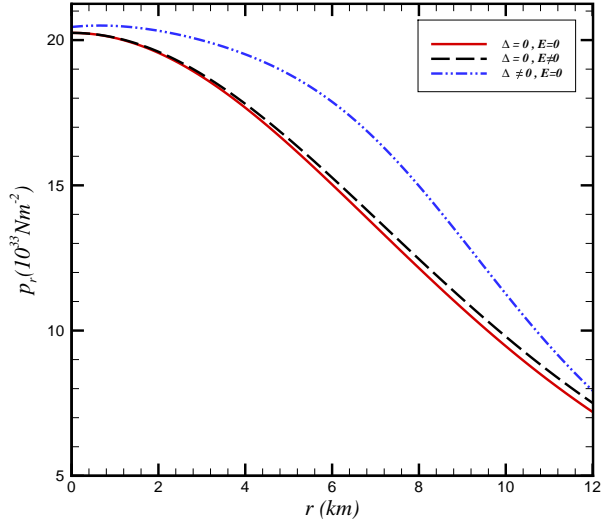} \\
\caption {Radial pressure $p_{r}$ for various cases: isotropic uncharged, isotropic charged, and anisotropic uncharged
 matter with $f=5\times10^{-5}esu$ $g^{-1}$, $A=3\times10^{33}Nm^{-2}$, $\sigma=3\times10^{3}m$ and $\mu = \frac{2R}{3}$ versus radial coordinate
$r$, all having the same central density of $\rho_{c}=1\times10^{18} kg$ $m^{-3}$.}
 \label{fig:p_rE}
\end{figure}

\begin{figure}
\centering
\includegraphics[scale=0.4]{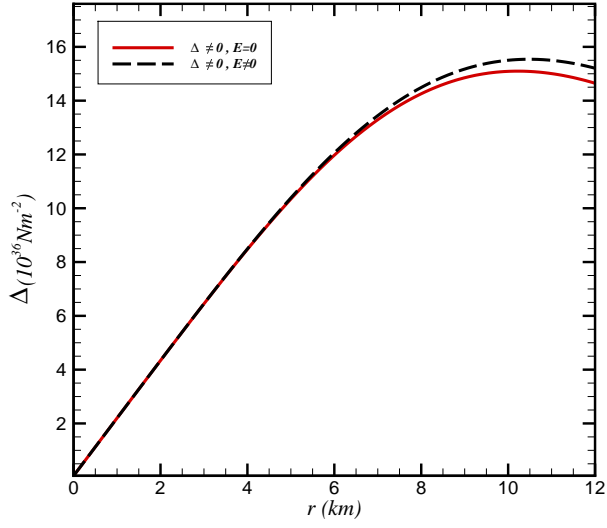} \\
\caption {Anisotropy factor $\Delta$ for anisotropic charged
and uncharged matter with $f=5\times10^{-5}esu$ $g^{-1}$ versus radial coordinate
$r$ with a same central density $\rho_{c}=1\times10^{18} kg$ $m^{-3}$.}
 \label{fig:deltaE}
\end{figure}

\begin{figure}
\centering
\includegraphics[scale=0.4]{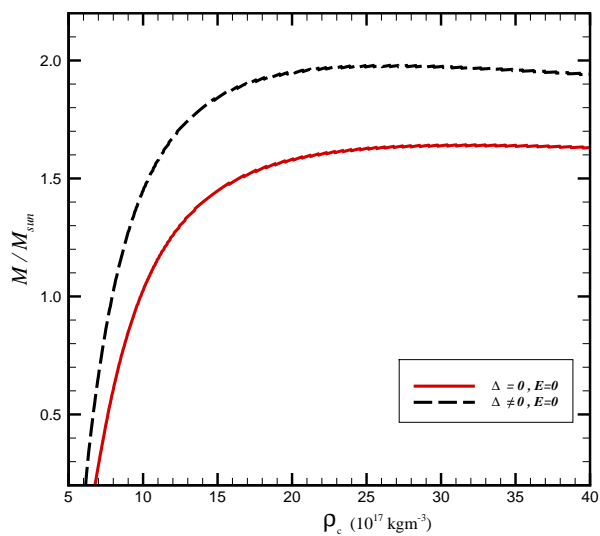} \\
\caption {Gravitational mass M vs central density $\rho_{c}$ for
uncharged isotropic (solid curve) and uncharged anisotropic (dashed curve) matter.}
 \label{fig:Mmax}
\end{figure}

\begin{figure}
\centering
\includegraphics[scale=0.4]{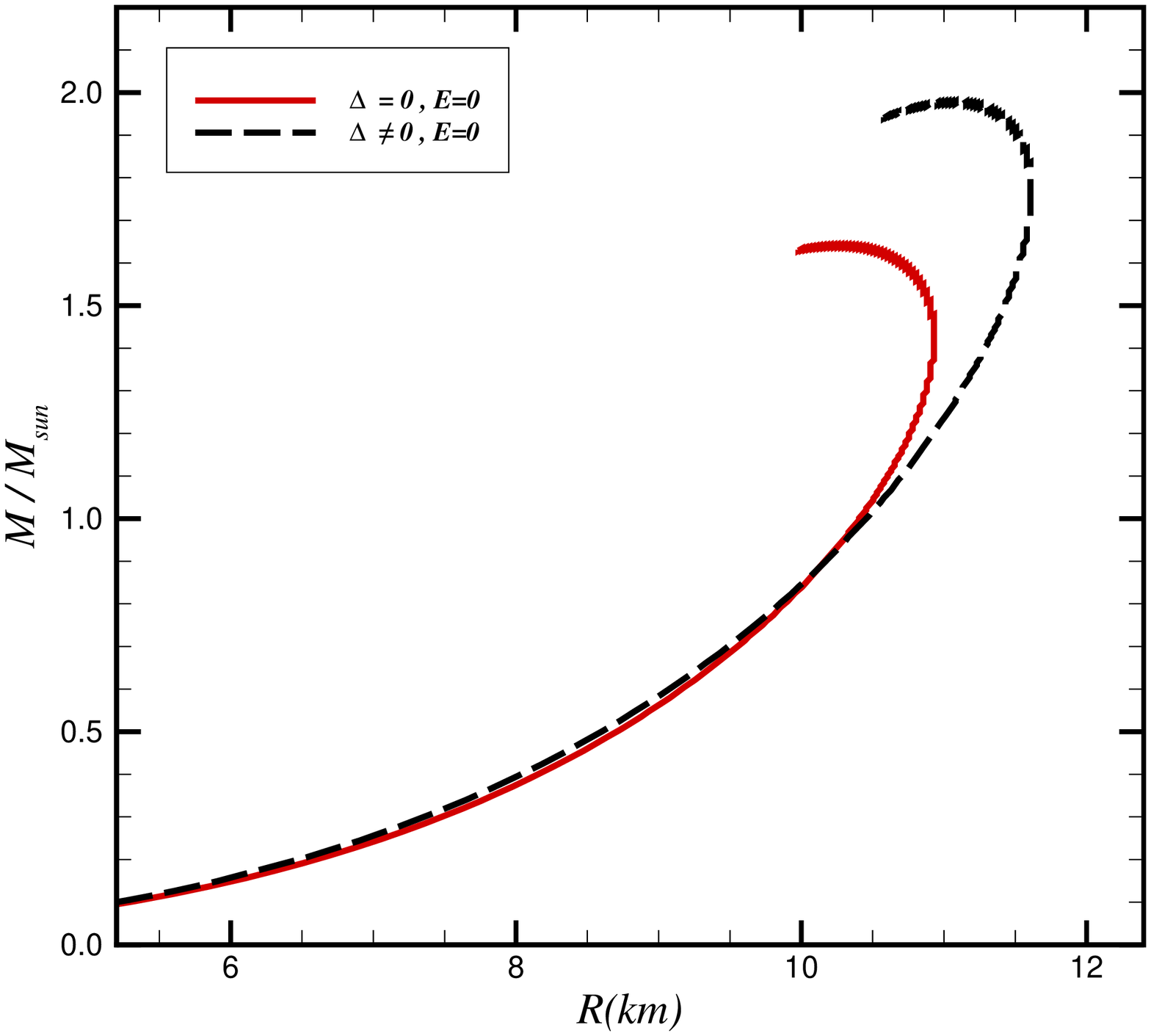} \\
\caption {Gravitational mass M vs radius of star R in two cases:
Uncharged isotropic (solid curve) and uncharged anisotropic (dashed curve) matter.}
 \label{fig:pich}
\end{figure}

It is shown in Fig. \ref{fig:Mmax} and Fig. \ref{fig:pich} that
the maximum mass of strange quark star has been raised by adding
Gaussian perturbation for obtaining $p_r$. Furthermore, according to Fig. \ref{fig:deltaE} the tangential pressure
$p_{t}$ and anisotropy $\Delta$ have a maximum and they do not
vanish on the surface. In fact our solution indicate that if there is an anisotropy in the star,
 it should be maximum on the surface. \\
Recent measurements indicate that there exist a  pulsar of  mass 1.97 $\pm$ 0.04 $M_{\odot}$ and this mass rules out nearly all currently proposed equations of state\cite{Demorest}. We have used a trial and error method to get  the mentioned  mass above. Our calculations indicate that an anisotropy amplitude of $A=3\times10^{33}Nm^{-2}$ with a standard deviation of $\sigma=3\times10^{3}m$ and $\mu = \frac{2R}{3}$ in Eq. (\ref{gaussian}) can survive SQM equation of state which satisfies boundary and hydrostatic equilibrium conditions by $1\%$ and $10\%$  uncertainties at the origin and on the surface respectively as is depicted in Fig. \ref{fig:p_rE}.

\section{Summary}
We have studied anisotropic charged strange quark stars with  a Gaussian perturbation. Our calculations have shown that the electrical charge effect is much less effective in increasing the maximum mass of star in comparison to anisotropy effect. Also our work has supported  that anisotropy can be one of the candidates of describing massive neutron stars with current equations of state.


\end{document}